\documentclass[aps,pre,twocolumn,groupedaddress,showpacs]{revtex4-1}

\usepackage{graphicx}
\usepackage{bm}
\usepackage{color}

\newcommand{\meanN}{\overline{N}}
\newcommand{\meanT}{\overline{T}}
\newcommand{\meanrho}{\overline{\rho}}
\newcommand{\meanIb}{\overline{I}_b}
\newcommand{\meanI}{\overline{I}}
\newcommand{\meankappa}{\overline{\kappa}}

\begin{document}

\title{Mechanism of unconfined dust explosions: turbulent clustering and radiation-induced ignition}
\author{Michael Liberman$^{1}$}
\email{mliber@nordita.org}
\author{Nathan Kleeorin$^{2,1}$}
\email{nat@bgu.ac.il}
\author{Igor Rogachevskii$^{2,1}$}
\email{gary@bgu.ac.il}
\author{Nils Erland L. Haugen$^{3,4}$}
\email{Nils.E.Haugen@sintef.no}

\medskip
\affiliation{
$^1$Nordita, KTH Royal Institute of Technology
and Stockholm University, Roslagstullsbacken 23,
10691 Stockholm, Sweden \\
$^2$Department of Mechanical Engineering,
Ben-Gurion University of the Negev, P. O. Box 653, Beer-Sheva
84105, Israel\\
$^3$SINTEF Energy Research, N-7034 Trondheim, Norway\\
$^4$Department of Energy and Process Engineering,
Norwegian University of Science and Technology,
Kolbj{\o}rn Hejes vei 1B, NO-7491 Trondheim, Norway
}

\date{\today}

\begin{abstract}
It is known that unconfined dust explosions typically
starts off with a relatively weak primary
flame followed by a severe secondary explosion.
We show that clustering of dust particles in a
temperature stratified turbulent flow ahead
of the primary flame may give rise to a significant
increase in the radiation penetration length.
These particle clusters, even far ahead of the flame,
are sufficiently exposed and heated by the radiation 
from the flame to become ignition kernels capable to ignite
a large volume of fuel-air mixtures.
This efficiently increases the total flame surface area
and the effective combustion speed, defined as the rate 
of reactant consumption of a given volume.
We show that this mechanism explains the high rate of combustion
and overpressures required to account
for the observed level of damage in unconfined dust explosions, e.g.,
at the 2005 Buncefield vapor-cloud explosion.
The effect of the strong increase of radiation transparency due to
turbulent clustering of particles goes beyond the state-of-the-art of the
application to dust explosions and has many implications in
atmospheric physics and astrophysics.
\end{abstract}


\maketitle

\section{Introduction}

As is known, dust explosions can occur when an accidentally ignited flame
propagates through a cloud of fine particles suspended in a combustible gas.
The danger of dust explosions have been known for centuries in the mining
industry and grain elevators, and it is currently a permanent threat in
all those industries in which powders of fine particles are involved.
A significant progress has been made in the development of prevention
and safety measures (see, e.g., \cite{Eckhoff2003,AmyotteEckhoff2010,Proust2006,AbbasiAbbasi2007,Yuan2015},
and references therein).
However, despite considerable efforts over more than 100 years,
the mechanism  of flame propagation in unconfined (large scale) dust explosions
still remains the main unresolved issue.

Based on the analysis of many
catastrophic accidents, it is currently  well established that
unconfined dust explosions consist of a relatively weak primary
explosion followed by a severe secondary explosion.
While the hazardous effect of the primary explosion is relatively small,
the secondary explosion may propagate with a speed of up to 1000 m/s,
producing overpressures of over 8-10 atm, which is comparable to the
pressures produced by a detonation.
However, the analysis of damages indicates
that a detonation is not involved, while normal deflagrations are
not capable of producing such high velocities and overpressures.
It is interesting to note that  M. Faraday and C. Lyell, who in 1845 analyzed the
Haswell coal mine explosion, where the first to point to
the likely key role of dust particles \cite{Faraday1845}.

Of special interest for understanding the nature of dust explosions is
the Buncefield vapor-cloud explosion,
for which a large amount of data and evidence were collected,
providing unique and valuable information about the timing
and damage caused by the event \cite{Buncefield2009}.
Later, a comprehensive analysis was performed by two groups
\cite{Atkinson-Cusco2011,Bradeley2012} in an attempt to
understand possible mechanisms leading to the unusually high rate of combustion.
Among different mechanisms, the hypothesis proposed by Moore and Weinberg
\cite{Moore81,Moore83}, where they argue that the anomalously high rate of
flame propagation in dust explosions can be due to the radiative
ignition of millimeter sized fibrous particles ahead of the flame front,
attracted much attention.

In normal practice, emissivity of combustion products and radiation absorption
in a fresh unburnt gaseous mixture are small and do not influence the
flame propagation.
The situation is drastically changed for  flames
propagating through a cloud of fine particles suspended in a gaseous mixture.
In this case, the radiative flux emitted from the primary flame
($\approx$ 300 - 400 kW/m$^{2}$), is close to the black-body radiation
\cite{Hadjipanayis2015}  at stoichiometric flame temperatures (2200 - 2500~K).
In dust explosions, the radiative flux into the reactants is significantly enhanced
by the increased emissivity of  the large volume of burned products.
Microns-size dust particles, with diameter $d_{p}$ that is
larger than the radiation wavelength $\lambda_{\rm rad}$,
efficiently absorb radiation emitted by the flame.
The absorbed heat is transferred to the surrounding gas by thermal
conduction, so that the gas temperature lags behind that of the particle.

If particles are evenly distributed, the maximum temperature increase
ahead of the flame is \cite{Liberman2015}:
\begin{eqnarray}
\Delta T_p \approx \frac{0.63 \, \sigma T_b^4}{U_{\rm f}(\rho_p
\, c_{\rm p} + \rho_g \, c_{\rm v})},
\label{AAA1}
\end{eqnarray}
where $U_{\rm f}$ is the normal flame velocity, $\sigma T_b^4$ is the
black-body radiative flux, and $\rho_p$, $\rho_g$, $c_{\rm p}$,
$c_{\rm v}$ are the mass density of particles, the gas mixture and
their specific heats, respectively.  As a flame propagates through a
dust cloud of evenly dispersed particles it consumes the unburned fuel
before the gas temperature will have risen up to ignition level, so
that radiation can not become a dominant process of the heat transfer
\cite{Liberman2015}.

Instead of being uniformly distributed, the particles may be organized in the
form of a distant optically thick layer or a clump ahead of the flame
front.
It was shown \cite{Liberman15} that if a transparent gap
between the particle layer and the flame is not too small, the
particle layer can be sufficiently heated by the flame-emanated
radiation to ignite new combustion modes in the surrounding fuel-air
mixture ahead of the main flame.

In turbulent flows ahead of the primary flame (${\rm Re}\approx 10^{4}$ - $10^{5}$),
typical for dust explosions, dust particles with
$\rho_p \gg \rho_g$ assemble in small clusters of
the order of the Kolmogorov turbulent scale,
$\ell_\eta \approx $0.1 - 1~mm.
The turbulent eddies, acting as small centrifuges, push the
particles to the regions between the eddies where the pressure
fluctuations are maximum and the vorticity intensity is minimum.
The particles assembled in these regions make clusters with 
higher particle number densities than the mean number density.
This effect, known as inertial clustering, has been investigated
in a number of analytical, numerical and experimental studies \cite{TB09,W09,BE10,EKR96,EKLR02,EKLR07,BB07,SA08,XB08,SSA08,SBB14}.

Recent analytical studies \cite{EKLR13,EKLR15} and laboratory experiments
\cite{EKR10} have shown that the particle clustering
can be much more effective in the presence of a mean temperature gradient.
In this case, the turbulence is temperature stratified,
and the turbulent heat flux is not zero.
This causes correlations between fluctuations of fluid temperature
and velocity, and, therefore, correlations between fluctuations
of pressure and fluid velocity, thus, producing additional pressure
fluctuations caused by the tangling of the mean
temperature gradient by the velocity fluctuations.
This enhance the particle clustering in the regions of maximum
pressure fluctuations.
As a result, the particle concentration in clusters rises
by a few orders of magnitude compared to the mean concentration
of evenly dispersed particles \cite{EKLR13}.

In order for a secondary dust explosion followed by a shock
wave to occur, the pressure from the mixture ahead of the flame
ignited by the radiatively heated particle clusters must
rise faster than it can be equalized by sound waves.
This means that the penetration length of radiation,
$L_{\rm rad}$, in the dust cloud
must be sufficiently large to ensure that the clusters of particles
even far ahead of the flame are sufficiently exposed and heated  by the
radiation, to become ignition kernels, i.e.,
$\tau_{\rm ign} \, c_{\rm s} \ll L_{\rm rad}$.
Here $c_{\rm s}$ is the sound speed in the mixture
ahead of the flame, and $\tau_{\rm ign}$  is the
characteristic time of fuel-air ignition by the
radiatively heated particle clusters (ignition kernels).

The effect of spatial inhomogeneities with the scales larger
than the wavelength of the radiation has been discussed in \cite{K93,R15b,K89}
and studied using Monte Carlo modeling \cite{Farbar16,Frankel16} of
the radiative heat transfer in particle-laden flow, taking into
account the inertial particle clustering in non-stratified turbulence.

In this paper we show that clustering of particles
ahead of the primary flame gives rise to a strong
increase of the radiation penetration (absorption) length.
The effect ensures that clusters of particles even far ahead of
the primary flame are sufficiently exposed and heated
by the radiation from the flame to become ignition kernels,
and the condition, $\tau_{\rm ign} \, c_{\rm s}
\ll L_{\rm rad}$, is satisfied.
The multiple radiation-induced ignitions in many ignition kernels ahead of the
primary flame increase the total flame surface area, so that the distance,
which each flame has to cover for a complete burn-out of the fuel, is
substantially reduced.
It results in a strong increase of the
effective combustion speed, defined as the rate of reactant
consumption of a given volume.
The proposed mechanism of unconfined dust explosions explains the
physics behind the secondary explosion.
It also explains the anomalously high rate of
combustion and overpressures required to account for the observed
level of damages in unconfined dust explosions.

\section{Mean-field equation for radiation transfer}

To describe this phenomenon, we consider a turbulent flow with suspended
particles that is exposed to a radiative flux.
The radiative transfer equation in the two-phase flow
\cite{R15,R15a} for the intensity of radiation,
$I({\bm r},\hat{\bm s})$ is
\begin{eqnarray}
&& \left(\hat{\bm s} {\bm \cdot} {\bm \nabla}\right)
I({\bm r},\hat{\bm s}) = - \left[\kappa_g({\bm r})
+ \kappa_p({\bm r})+\kappa_s({\bm r})\right]\, I({\bm r},\hat{\bm s})
\nonumber\\
&& \; \; + \kappa_g \, I_{b,g} + \kappa_p I_{b,p} + {\kappa_s
\over 4 \pi} \int \phi({\bm r},\hat{\bm s},\hat{\bm s}')
\, I({\bm r},\hat{\bm s}') \,d\Omega ,
\label{A1}
\end{eqnarray}
where $\kappa_g({\bm r})$ and $\kappa_p({\bm r})$ are the absorption coefficients
of gas and particles, respectively, $\kappa_s({\bm r})$ is the particle
scattering coefficient, $\phi$ is the scattering phase function,
$I_{b,g}({\bm r})$ and $I_{b,p}({\bm r})$ are the black-body emission intensities
for gas and particles, respectively, which depend on the local
temperature, $\hat{\bm s}={\bm k}/k$ is the unit vector in the
direction of radiation, ${\bm k}$ is the wave vector,
${\bm r}$ is the position vector,
$\,d\Omega=\sin \theta \,d\theta \,d\varphi$, and
$(k, \theta, \varphi)$ are the spherical coordinates in ${\bm k}$ space.
Since the scattering and absorption cross sections for gases at normal
conditions are very small, the contribution from the gas phase can be
neglected in comparison with that of particles.
Furthermore, we assume that the particle absorption coefficient is
much larger than the scattering coefficient,
$|\kappa_p({\bm r})| \gg | \kappa_s({\bm r})|$,
which is the case when $\pi d_p \sqrt{|\epsilon|}/\lambda_{\rm rad} \gg 1$ \cite{N82},
where $\epsilon$ is the dielectric permeability of the dust particles,
$\lambda_{\rm rad}$ is the radiation wavelength and $d_{p}$ is the diameter
of the dust particles.
Under these assumptions Eq.~(\ref{A1}) is reduced to
\begin{eqnarray}
&& \left(\hat{\bm s} {\bm \cdot} {\bm \nabla}\right)
I({\bm r},\hat{\bm s}) = - \kappa({\bm r}) \left(I({\bm r}) - I_b\right) ,
\label{A2}
\end{eqnarray}
where we denote $\kappa \equiv\kappa_p({\bm r})$ and $I_b\equiv I_{b,p}$.

The radiation absorption length of evenly dispersed particles is
$L_a=1/\meankappa$, where $\meankappa = \sigma_a \meanN$ is the mean particle absorption
coefficient, $\sigma_a\approx\pi d_p^2/4$ is the particle absorption
cross section, and  $\meanN$ is the mean number density of evenly
dispersed particles.
For a typical mean dust density
$\meanrho_d \equiv \meanN \, m_{\rm p}=$0.1-0.3 kg/m$^3$,
the corresponding radiation absorption length is
${L_a} = 1/\meankappa \approx (\rho_p/2\rho_d)\, d_p \approx$ 10 cm,
where ${\rho_p} = 1$ kg/cm$^3$ and $d_p= 10 \, \mu$m.
If particles are accumulated in optically thick clusters, with
the particle number density within clusters, $n_{\rm cl}$, much
larger than $\meanN$, the effective radiation absorption length can be
roughly estimated as $L_{\rm eff} = 1 / (\sigma_{\rm cl} N_{\rm cl})$,
where $\sigma_{\rm cl}$ is the cluster cross-section and $N_{\rm cl}$
is the mean number density of clusters.
Taking into account that the cluster size is of the
order of the Kolmogorov turbulent scale $\ell_\eta$,
we obtain: $L_{\rm eff} / L_a \approx (n_{\rm cl} / \meanN) \, (\ell_\eta / L_a)$.
This expression shows that the result depends on the interplay
between a large factor $n_{\rm cl} / \meanN \gg 1$ and a small factor
$\ell_\eta/L_a \ll 1$, which in turn depend on the parameters of turbulence.

To determine the effective
radiation absorption coefficient for a turbulent flow with
particle clustering, we apply a mean-field approach, i.e.,
all quantities are decomposed into the mean
and fluctuating parts: $I=\meanI + I'$, $I_b=\meanIb + I'_b$
and $\kappa=\meankappa + \kappa'$, where the fluctuating parts,
$I', I'_b, \kappa'$  have zero mean values, and overbars
denote averaging over an ensemble.
The instantaneous particle absorption coefficient is
$\kappa=n \, \sigma_a$, and the mean particle absorption
coefficient is $\meankappa=\meanN \, \sigma_a$, so that
the fluctuations of the particle absorption coefficient
are $\kappa' = n' \, \sigma_a= n' \, \meankappa/\meanN$.
Here $n=\meanN+n'$ is the instantaneous particle number
density and $n'$ are fluctuations of the particle number density.

We average Eq.~(\ref{A2}) over the ensemble of the particle
number density fluctuations and assume that the characteristic
correlation scales of fluctuations are much smaller than the
scales of the mean field variations.
The equation for the mean radiation intensity
$\meanI({\bm r},\hat{\bm s})$ reads:
\begin{eqnarray}
&& \left(\hat{\bm s} {\bm \cdot} {\bm \nabla}\right)
\meanI({\bm r},\hat{\bm s}) = - \meankappa
\left(\meanI - \meanIb\right) - \langle \kappa' \, I' \rangle
+ \langle \kappa' \, I'_b \rangle ,
\label{A3}
\end{eqnarray}
where the angular brackets $\langle ... \rangle$ denote
averaging over an ensemble of fluctuations.
To determine the one-point correlation function
$\langle \kappa' \, I' \rangle$,
we need to derive equations for fluctuations $I'$ and $\kappa'$.
By subtracting Eq.~(\ref{A3}) from Eq.~(\ref{A2})
we obtain the equation for fluctuations of $I'$:
\begin{eqnarray}
&& \left(\hat{\bm s} {\bm \cdot} {\bm \nabla} + \meankappa
+ \kappa' \right) I'({\bm r},\hat{\bm s}) = I_{\rm source} ,
\label{A5}
\end{eqnarray}
where the source term is
\begin{eqnarray}
&& I_{\rm source}= - \kappa' \, \left(\meanI - \meanIb\right)
+ \langle \kappa' \, I' \rangle +
\left(\meankappa + \kappa'\right)I'_b
 - \langle \kappa' \, I'_b \rangle  .
\nonumber\\
\label{AA5}
\end{eqnarray}
The solution of Eq.~(\ref{A5}) reads:
\begin{eqnarray}
&& I'({\bm r},\hat{\bm s}) = \int_{-\infty}^{\infty}
I_{\rm source} \, \exp\left[-\left|\int_{s'}^{s}
\left[\meankappa+ \kappa'(s'') \right] \,ds'' \right| \right]\,ds' ,
\nonumber\\
\label{A6}
\end{eqnarray}
where $s={\bm r} {\bm \cdot} \hat{\bm s}$.
Expanding the function $\exp[-|\int_{s'}^{s} \kappa'(s'')| \,ds'']$
in Taylor series, multiplying Eq.~(\ref{A6}) by $\kappa'$
and averaging over the ensemble of fluctuations,
we obtain the following
equation for the one-point correlation function
$\langle \kappa' \, I' \rangle$:
\begin{eqnarray}
\langle \kappa' \, I' \rangle  = - \meankappa
\left(\meanI - \meanIb\right) \, {2 \beta J_1 \over 1 + 2 \beta J_2} ,
\label{A7}
\end{eqnarray}
where $\beta=\meankappa \, \ell_D$ and
$\ell_D =a \, \ell_\eta / {\rm Sc}^{1/2}$
with a numerical coefficient $a \gg 1$, while the Schmidt number 
is given by ${\rm Sc}=\nu/D$, $\nu$ is the kinematic viscosity,
$D$ is the coefficient of the Brownian diffusion of particles,
\begin{eqnarray}
&& J_1= \int_{0}^{\infty} \Phi(Z) \exp(-\beta Z) \,dZ ,
\label{A8}\\
&& J_2= \beta \int_{0}^{\infty} \left(\int_{0}^{Z} \Phi(Z') \,dZ' \right)
\, \exp(-\beta Z) \,dZ ,
\label{A9}
\end{eqnarray}
and $\Phi({\bf R}) = \langle n'({\bf x}) \, n'({\bf x} + {\bf R})\rangle$
is the two-point instantaneous correlation function of the
particle number density fluctuations.
In derivation of Eq.~(\ref{A7}) we applied the quasi-linear
approach, i.e., we neglected the third and higher
moments in fluctuations of $\kappa'$ in Eq.~(\ref{A7})
due to a small parameter $\kappa' \ell_\eta {\rm Sc}^{-1/2} \ll 1$.
We also assumed that $\langle \kappa' \, I'_b \rangle = 0$, i.e.,
we neglected the correlation between the particle
number density fluctuations and the fluid temperature fluctuations.

Substituting Eq.~(\ref{A7}) for $\langle \kappa' \, I' \rangle$
into Eq.~(\ref{A3}) we obtain the mean-field
equation for the mean radiation intensity:
\begin{eqnarray}
\left(\hat{\bm s} {\bm \cdot} {\bm \nabla}\right)
\meanI({\bm r},\hat{\bm s}) = - \kappa_{\rm eff} \,
\left(\meanI - \meanIb\right) ,
\label{A12}
\end{eqnarray}
where the effective absorption coefficient, $\kappa_{\rm eff}$, is
\begin{eqnarray}
\kappa_{\rm eff} = \meankappa \, \left(1 - {2 \beta J_1 \over 1 + 2 \beta J_2}\right) .
\label{A14}
\end{eqnarray}
The last term on the right hand side of Eq.~(\ref{A14})
takes into account the particle clustering in stratified turbulence.

\section{Particle correlation function}

To calculate the integrals $J_1$ and $J_2$,
we use the normalized two-point correlation function of the
particle number density fluctuations $\Phi(R)=\langle n'({\bm r}) \,
n'({\bm r}+{\bm R}) \rangle / \meanN^2$, derived in \citep{EKLR13}
for a temperature stratified turbulence:
$\Phi(R) = (n_{\rm cl} / \meanN)^2$, for $0\leq R\leq \ell_D$,
\begin{eqnarray}
\Phi(R) = \left({n_{\rm cl} \over \meanN}\right)^2 \,
\left({R \over \ell_D}\right)^{-\mu} \cos\left[\alpha \,
\ln \left({R \over \ell_D}\right)\right],
 \label{A10}
\end{eqnarray}
for $\ell_D \leq R\leq \infty$,
where $R={\bm R} {\bm \cdot} \hat{\bm s}$ and
\begin{eqnarray}
\mu &=& {1 \over 2(1+3\sigma_{_{T}})} \left[3 - \sigma_{_{T}} + {20 \sigma_v (1+\sigma_{_{T}})\over 1+\sigma_v}\right],
\label{C1}\\
\alpha &=& {3\pi (1+\sigma_{_{T}})\over (1+3\sigma_{_{T}}) \, \ln {\rm Sc}} .
\label{CC1}
\end{eqnarray}
Here $\sigma_{_{T}} = \left(\sigma_{_{T0}}^2 + \sigma_v^2\right)^{1/2}$
is the degree of compressibility of the turbulent diffusion tensor and
$\sigma_v$ is the degree of compressibility of the particle velocity field:
\begin{eqnarray}
\sigma_v &=& {8 {\rm St}^2 \, K^2 \over 3(1 + b \, {\rm St}^2 \, K^2)} ,
\; \; K = \left[1+ {\rm Re} \,\left({L_\ast \, {\bm \nabla} \meanT \over \meanT}
\right)^2\right]^{1/2} ,
\nonumber\\
 \label{C2}
\end{eqnarray}
${\rm Re} = \ell_0 u_0 /\nu$ is the Reynolds number,
$\sigma_{_{T0}} = \sigma_{_{T}}({\rm St}=0)$,
and a constant $b$ determines the value of the parameter
$\sigma_v$ in the limit of ${\rm St}^2 \, K^2 \gg 1$.
The ratio of the particle Stokes
time, $\tau_p=m_p/(3 \pi \rho \, \nu d_p)$,
and the Kolmogorov time,
$\tau_\eta= \tau_0 /{\rm Re}^{1/2}$, is the Stokes number,
${\rm St} =\tau_p/{\tau}_\eta$,
where $\tau_0=\ell_0/u_0$ and $u_0$
are the characteristic time and velocity at the turbulent
integral scale, $\ell_0$, respectively.
In Eq.~(\ref{C2}), $\meanT$ is the mean fluid temperature and
the length $L_\ast = c_{\rm s}^2 \tau_\eta^{3/2}/9 \nu^{1/2}$.

To derive Eqs.~(\ref{A10})--(\ref{C2}) we
used Eqs.~(42)-(47) in  \citep{EKLR13},
and Eqs.~(17)-(18) in \citep{EKLR15}.
The correlation function $\Phi(R)$ takes into account particle
clustering.
For the inertial clustering in a non-stratified turbulence
$K=1$, while for the temperature stratified turbulence
the non-dimensional function $K$ is determined by Eq.~(\ref{C2}).

\section{Effective radiation length}

\begin{figure}
\vspace*{1mm} \centering
\includegraphics[width=8.0cm]{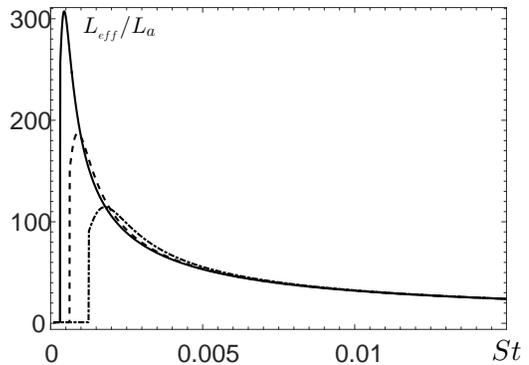}
\caption{\label{Fig1}
The ratio $L_{\rm eff}/L_{a}$ versus the Stokes number ${\rm St}$ for different mean temperature gradients $|{\bm \nabla} \meanT|$: 0.225 K/m (dashed-dotted), 0.45 K/m (dashed), 0.9 K/m (solid).}
\end{figure}

\begin{figure}
\vspace*{1mm} \centering
\includegraphics[width=8.0cm]{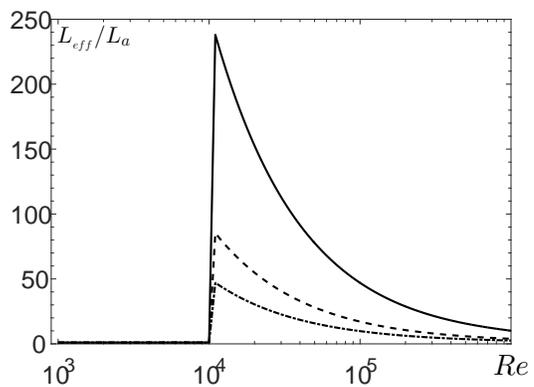}
\caption{\label{Fig2}
The ratio $L_{\rm eff}/L_{a}$ versus the Reynolds number ${\rm Re}$ for different particle diameter $d_p$: 3 $\mu$m (solid), 6 $\mu$m (dashed), 9 $\mu$m (dashed-dotted),
and for the temperature gradients $|{\bm \nabla} \meanT|= 0.9$ K/m.}
\end{figure}

Integrals $J_1$ and $J_2$ are calculated using Eqs.~(\ref{A10})--(\ref{C2}) and taking into account that $\beta=\meankappa \, \ell_D \ll 1$,
\begin{eqnarray}
J_1=J_2 &=& \left({n_{\rm cl} \over
\meanN}\right)^2 \, \left[1 + {\mu-1 \over (\mu-1)^2 + \alpha^2}\right] .
\label{A11}
\end{eqnarray}
Substituting Eqs.~(\ref{A11}) into Eq.~(\ref{A14}), we determine the effective penetration
length of radiation $L_{\rm eff} \equiv 1/\kappa_{\rm eff}$ as:
\begin{eqnarray}
{L_{\rm eff} \over L_a} =1 +
{2 a \over {\rm Sc}^{1/2}} \, \left({n_{\rm cl} \over \meanN}\right)^{2} \, \left({\ell_\eta \over L_a}\right)\, \left[1 + {\mu-1 \over (\mu-1)^2 + \alpha^2}\right]  .
\nonumber\\
\label{A16}
\end{eqnarray}
Figure~\ref{Fig1} shows the ratio $L_{\rm eff}/L_{a}$
versus the Stokes number ${\rm St}$ for different temperature
gradients $|{\bm \nabla} \meanT|$, calculated for parameters typical
for unconfined dust explosions:  ${\rm Re}=5 \times 10^4$, $\ell_0=1$ m, $u_0=1$ m/s;
$\nu=0.2$ cm/s$^2$, $c_{\rm s}=450$ m/s for methane-air,
$n_{\rm cl} /\meanN = 500$ and $\sigma_{_{T0}} =1/2$.
Using asymptotic analysis we choose $a=10$ and $b=5$.
Figure~\ref{Fig2} shows the ratio $L_{\rm eff}/L_{a}$
versus the Reynolds numbers for particles of different diameters.

\section{Discussion}

It is shown that  clustering of particles in the temperature
stratified turbulent flow ahead of the primary flame gives rise to a strong increase of the radiation penetration length (Fig.~\ref{Fig1}), but within a narrow interval of turbulent parameters (see Fig.~\ref{Fig2}).
This explains the mechanism of the secondary explosion in unconfined dust explosions.
According to analysis of the Buncefield explosion \cite{Atkinson-Cusco2011}:
``The high overpressures in the cloud and low average rate of flame
advance can be reconciled if the rate of flame advance was episodic,
with periods of very rapid combustion being punctuated by pauses when
the flame advanced very slowly.'' Assuming $L_{a}\sim 10$ cm, we
obtain $L_{\rm eff}\sim 10$ m as the result of the turbulent
clustering of dust particles.
Given that the radiative flux from the primary
flame is about $S\sim$ 300 - 400 kW/m$^2$, the temperature of
particles increases up to the ignition level $\Delta T_p\approx
1000$~K during ${\tau}_{\rm ign}= (\rho_p \, d_p \, c_{\rm p} \,
\Delta T_p)/2S < 10$ ms.
Time-scales of the fuel-air ignition by the
radiatively heated cluster of particles (with $d_p$ in the range from 
1 to $20 \, \mu$m) can be smaller than 10 ms \cite{Liberman2015,R7}.
This ensures that the condition $L_{\rm eff} \gg {c_{\rm s}} \, \tau_{\rm ign}$ is
satisfied, and the pressure produced by the ignition of the fuel-air mixture
rises until the pressure wave steepen into a shock wave.
The effective rate of the secondary explosion propagation is $\sim 1000$
m/s, which corresponds to a Mach number, ${\rm Ma}=$2.5 - 3, and
overpressures of 8-10 atm.
This explains the level of damage observed
in the aftermath of unconfined dust explosions
\cite{Buncefield2009,Atkinson-Cusco2011}.
At the same time, the secondary explosion strongly change
parameters of the turbulent flow.
As a result, rapid combustion in secondary explosion is interrupted
until the shock wave run away, and the slow combustion parameters of
the turbulent flow ahead of a slow deflagration wave 
will be suitable for formation of a new transparent radiation window.
This agrees fairly well with the Buncefield explosion
scenario \cite{Atkinson-Cusco2011}.
The effect of the strong increase of the radiation penetration length due to
turbulent particle clustering goes beyond to the dust explosion applications
and has many implications in astrophysical and atmospheric turbulence
\cite{OS16,SOC12,EKLR15}.

\begin{acknowledgements}
This work has been supported by the Research Council of Norway
under the FRINATEK (grant  No. 231444).
The support of this work provided by the opening project number 
KFJJ17-08M of State Key Laboratory of Explosion Science and 
Technology, Beijing Institute of Technology is gratefully acknowledged (ML).
Stimulating discussions with participants of the NORDITA program
on ``Physics of Turbulent Combustion'' are acknowledged.
\end{acknowledgements}

\end{document}